\documentclass[conference]{IEEEtran}
\IEEEoverridecommandlockouts
\usepackage{cite}
\usepackage{amsmath,amssymb,amsfonts}
\usepackage{algorithmic}
\usepackage{graphicx}
\usepackage{textcomp}
\usepackage{xcolor}
\usepackage{booktabs} 
\usepackage{multirow} 
\usepackage{subfig}	  
\def\BibTeX{{\rm B\kern-.05em{\sc i\kern-.025em b}\kern-.08em
    T\kern-.1667em\lower.7ex\hbox{E}\kern-.125emX}}
\begin{document}

\title{SI-FID: Noise-Aware Fine-Tuning for Perceptual Quality Assessment of Stitched Images}

\author{Shengwei~guo, Guobing~Sun}

\maketitle

\begin{abstract}
Accurate evaluation of stitched image quality is essential for advancing stitching algorithms, yet existing objective metrics often diverge from human perception because they insufficiently capture stitching-specific artifacts such as ghosting and misalignment. To address this limitation, we propose SI-FID, a noise-aware extension of the Fréchet Inception Distance tailored for stitched-image assessment. Instead of modifying the FID formulation itself, SI-FID adapts the underlying feature representation through contrastive fine-tuning with controlled perturbations introduced via data augmentation, thereby enhancing sensitivity to subtle stitching-induced distortions. A pre-trained InceptionV3 encoder is calibrated using both original and perturbed samples, yielding a perceptually aligned feature space for distribution-based quality evaluation. Experiments on two complementary benchmark datasets demonstrate that SI-FID improves rank correlation with human subjective scores by over 25\% relative to conventional metrics, providing a more reliable and perceptually consistent indicator for stitched image quality.
\end{abstract}


\section{Introduction}

Subjective evaluation, typically based on mean opinion scores (MOS), remains the most reliable form of quality assessment, but it is expensive, labor-intensive, and impractical for large-scale stitched-image testing~\cite{park2023probabilistic,madhusudana2022image,tang2024clip}. Objective IQA methods—including full-reference (FR) metrics such as MSE, PSNR, SSIM, and FSIM, and no-reference (NR) metrics such as BRISQUE and NIQE—estimate perceptual quality using luminance, structural, or statistical cues. While effective for generic distortions, these metrics are not specifically designed to emphasize the subtle local geometric inconsistencies characteristic of image stitching.

Feature-distance metrics, particularly the Fréchet Inception Distance (FID)~\cite{FID}, measure distributional differences between deep feature embeddings and correlate well with human judgments in generative modeling~\cite{zhou2023quality,duan2023attentive,madhusudana2022image,aslam2023vrl,golestaneh2022no,xiang2021no,kynkaanniemi2019improved}. By modeling global feature distribution alignment, FID provides a higher-level perceptual measure compared with pixel-based metrics. However, standard FID relies on features trained for image classification and is primarily sensitive to global semantic statistics, making it less responsive to localized stitching distortions.

\begin{figure}[!t]
	\centering
	\subfloat[]{\includegraphics[width=3.5in]{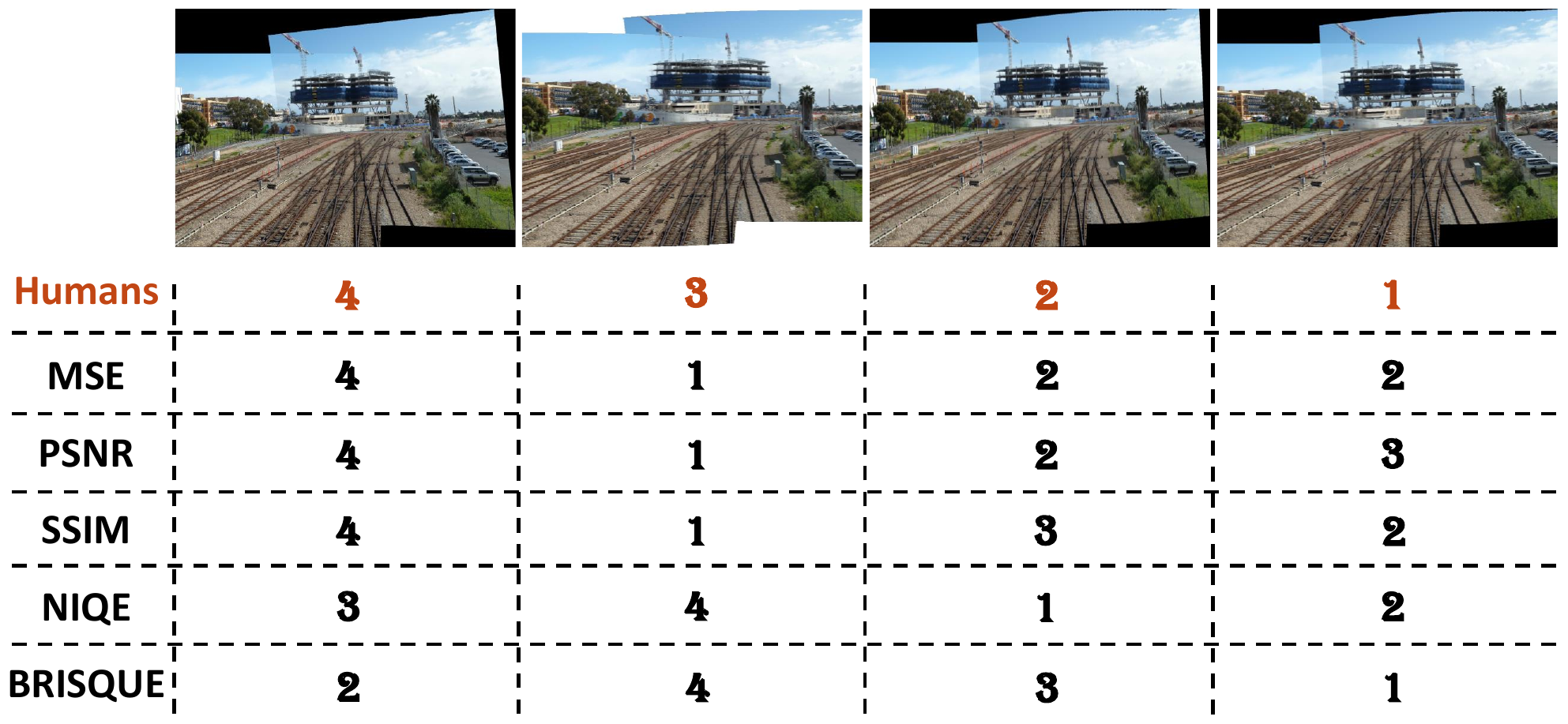}
	\label{fig_1_1}}
	\hfil
	\subfloat[]{\includegraphics[width=3.5in]{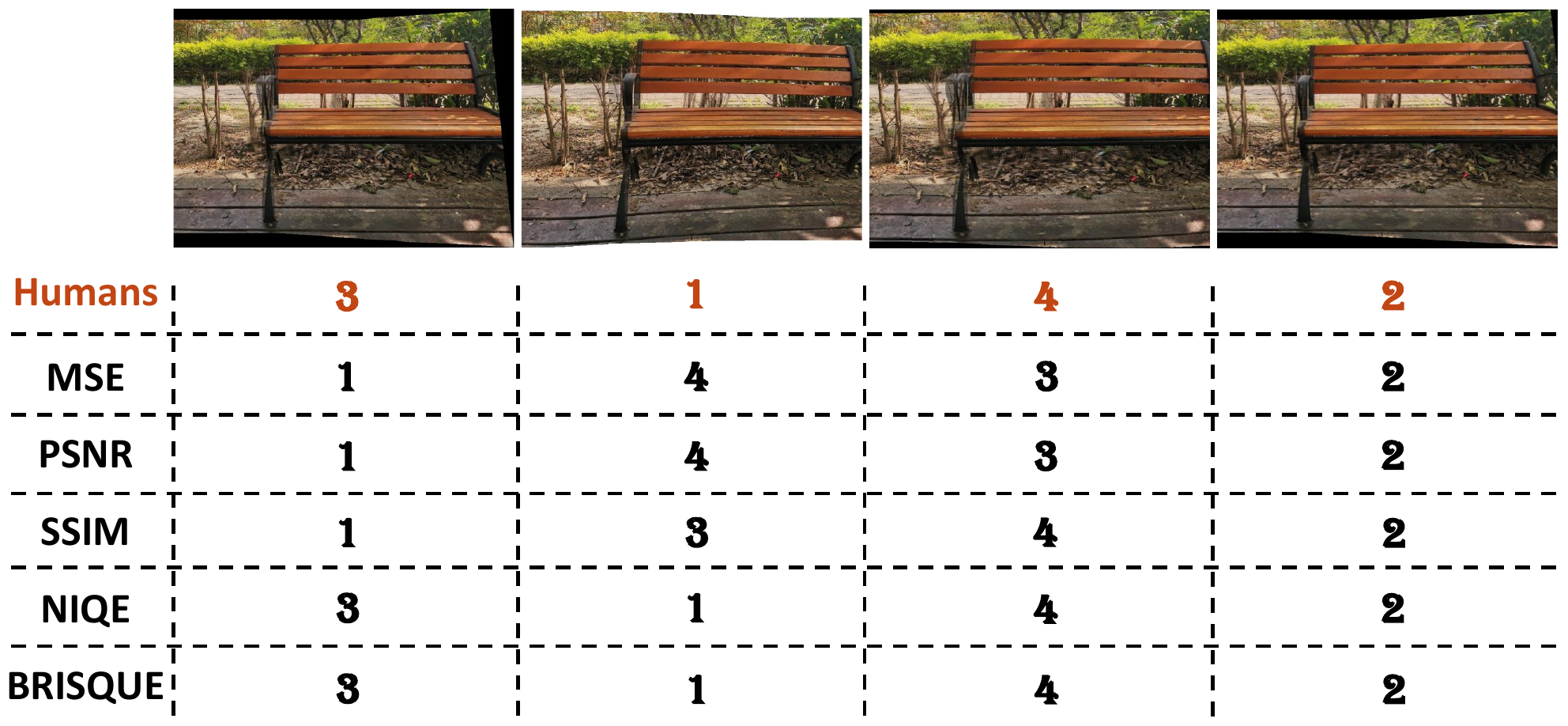}%
	\label{fig_1_2}}
	\caption{Examples illustrating the mismatch between subjective and objective assessments. Each pair shows stitched results from the same source images. Human scores are averaged from 14 participants, who consistently preferred one result, whereas common metrics (MSE, PSNR, SSIM, NIQE, BRISQUE) often produce conflicting or reversed rankings. Higher scores indicate better quality.}
	\label{fig:human_vs_pc}
\end{figure}

Existing stitched-image quality evaluation (SIQA) methods~\cite{10,9,11,12} introduce seam-based or task-specific cues to detect local artifacts, yet these approaches are often tailored to particular distortion types and may lack robustness across diverse scenes. To our knowledge, there is limited work that explicitly adapts feature-distribution metrics to enhance sensitivity to stitching-induced distortions through task-aware representation calibration.

To address these limitations, we propose the \emph{Stitched Image Fréchet Inception Distance (SI-FID)}, a noise-aware extension of FID. Rather than modifying the FID formulation itself, SI-FID adapts the underlying feature embedding space through contrastive fine-tuning guided by controlled perturbations. By reformulating stitched-image assessment as a task-aware feature-distance learning problem, the proposed approach enhances representation sensitivity to stitching artifacts while preserving global semantic consistency.
Extensive experiments on two stitched-image datasets demonstrate that SI-FID improves rank correlation with human subjective evaluation by over 25\% compared with classical FR, NR, and feature-based metrics. The main contributions are summarized as follows:

\begin{figure*}[!t]
	\centering
	\includegraphics[width=0.75\textwidth]{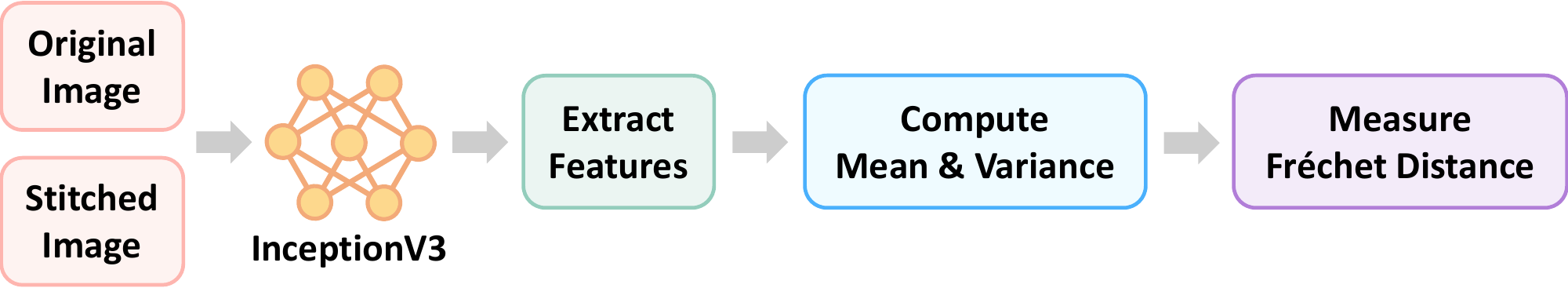}
	\caption{Overview of the standard FID-based image quality assessment pipeline. Features extracted by a pre-trained InceptionV3 model are modeled as Gaussian distributions, and their similarity is measured using the Fréchet distance. SI-FID extends this pipeline by introducing noise-aware fine-tuning to improve sensitivity to stitching-specific distortions.}
	\label{fig:si_fid_over}
\end{figure*}

\begin{itemize}
\item We reformulate stitched image quality assessment as a task-aware feature distribution calibration problem and introduce SI-FID, a noise-aware variant of FID.
\item We design a contrastive fine-tuning scheme with controlled perturbations to enhance feature sensitivity to stitching-induced distortions.
\item Experiments on two datasets show that SI-FID achieves substantially higher PCC/SROCC correlation and improved robustness compared with existing objective indicators~\cite{PCC,SROCC}.
\end{itemize}

\begin{figure}[h]
	\centering
	\includegraphics[width=0.35\textwidth]{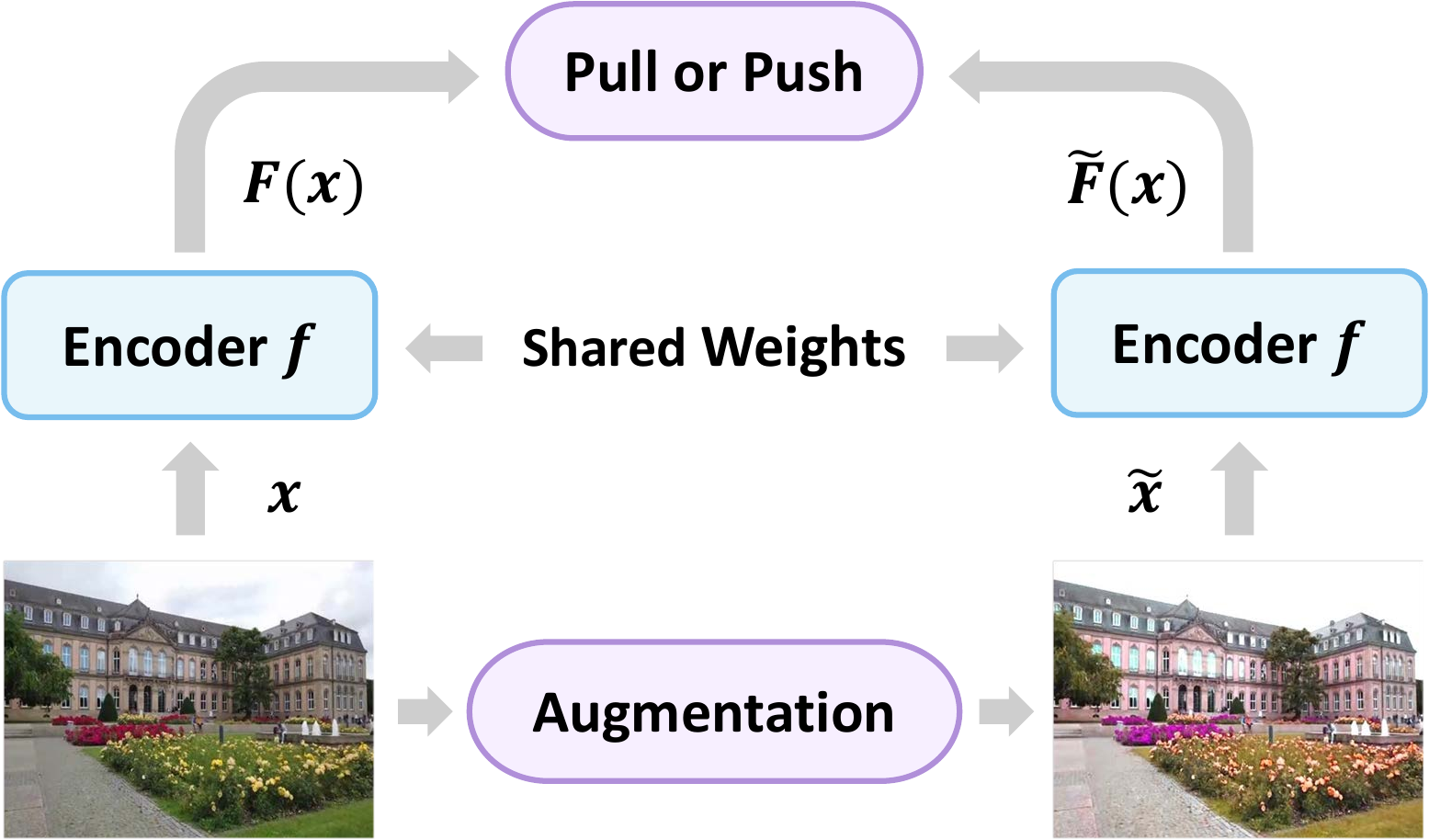}
	\caption{Proposed contrastive fine-tuning framework for SI-FID. Each image and its augmented counterpart are encoded by a shared InceptionV3 backbone. The cosine similarity loss encourages perceptually consistent embeddings while enhancing sensitivity to stitching distortions.}
	\label{fig_4}
\end{figure}

\begin{figure}[h]
	\centering
	\subfloat[]{\includegraphics[width=3.2in]{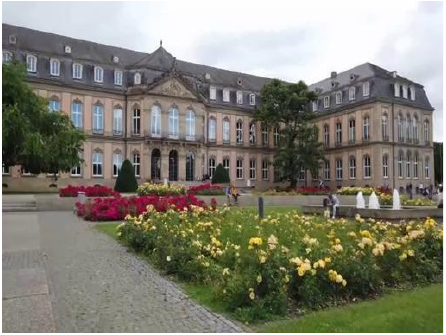}
		\label{fig_5a}}
	\hfil
	
	\subfloat[]{\includegraphics[width=3.2in]{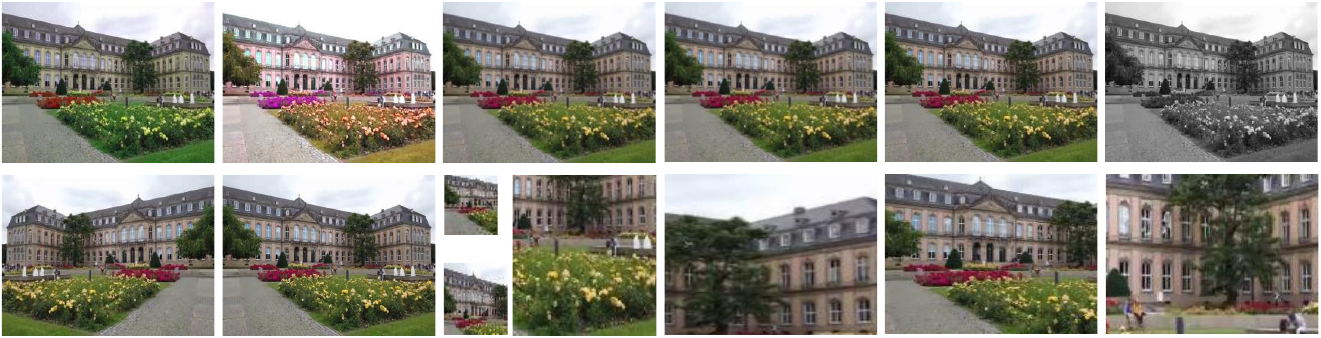}
		\label{fig_5b}}
	\caption{
		Examples of data augmentation used for SI-FID fine-tuning. 
		(a) Original image. 
		(b) Twelve variants generated via blur, cropping, rotation, and color jittering to simulate stitching artifacts such as misalignment and ghosting.}
	\label{fig_5}
\end{figure}

\section{Method}

This section presents the proposed Stitched Image Fréchet Inception Distance (SI-FID), which enhances FID by adapting the feature embedding space through controlled perturbation-driven fine-tuning, thereby improving sensitivity to stitching artifacts.

\subsection{Motivation}

Although FID captures global feature statistics effectively, it remains relatively insensitive to localized stitching distortions such as ghosting and misalignment. These artifacts typically occur in confined spatial regions and may not significantly alter global feature distributions, leading to weak correlation with human perceptual judgment.

Instead of modifying the FID formulation itself, we reinterpret FID as a feature-distribution metric whose perceptual behavior is determined by the underlying encoder. By calibrating the encoder to become more responsive to stitching-induced inconsistencies, the resulting distributional distance becomes more perceptually aligned. Inspired by robustness-oriented strategies such as NoisyTune~\cite{18}, we introduce controlled perturbations at the data level rather than injecting noise into model parameters. This encourages the encoder to amplify stitching-sensitive discrepancies while maintaining stability on clean inputs.

The overall framework is illustrated in Fig.~\ref{fig:si_fid_over}, where the standard FID pipeline serves as the backbone and the proposed noise-aware contrastive calibration is applied to the feature extractor.

\subsection{Fréchet Inception Distance}

Given feature embeddings of source images $I_s$ and stitched images $I_t$ extracted by InceptionV3~\cite{15}, FID models their feature distributions as multivariate Gaussians $\mathcal{N}(\mu_s,\Sigma_s)$ and $\mathcal{N}(\mu_t,\Sigma_t)$. The Fréchet Inception Distance (FID) is defined as:

\begin{equation}
\mathrm{FID} 
= \|\mu_s - \mu_t\|_2^2 
+ \mathrm{Tr}\!\left( \Sigma_s + \Sigma_t - 2(\Sigma_s \Sigma_t)^{1/2} \right),
\end{equation}

where $\|\cdot\|_2$ denotes the $\ell_2$ norm and $\mathrm{Tr}(\cdot)$ denotes the trace operator. While this formulation measures global distributional similarity, its perceptual sensitivity depends entirely on the learned feature representation. This motivates adapting the encoder to better encode stitching-specific distortions.

\subsection{Contrastive Fine-tuning} \label{subsubsec3}

To enhance stitching sensitivity, we fine-tune the InceptionV3 encoder $f(\cdot)$ within a contrastive learning framework~\cite{SimSiam}. For each training sample $x$ and its perturbed version $\tilde{x}$, the encoder produces feature embeddings $F=f(x)$ and $\tilde{F}=f(\tilde{x})$. The cosine similarity objective is defined as

\begin{equation}
L = - \frac{1}{2}\left[
\frac{F}{\|F\|_2} \cdot \frac{\tilde{F}}{\|\tilde{F}\|_2}
\right],
\end{equation}

which encourages semantically consistent pairs to remain close in the embedding space while increasing representation sensitivity to perturbation-induced structural inconsistencies. Through this process, the encoder learns to emphasize subtle distortions that are characteristic of stitching artifacts.

\subsection{Noise-Perturbation and Data Augmentation}

To approximate stitching-induced inconsistencies, we apply controlled perturbations via data augmentation, including Gaussian blur, random resized cropping, grayscale conversion, color jittering, and horizontal flipping. Each transformation is evaluated under multiple strengths to explore different distortion sensitivities. 

Moderate perturbations preserve global scene semantics while introducing localized inconsistencies, enabling the encoder to develop distortion-aware feature representations. Representative augmented examples are shown in Fig.~\ref{fig_5}.

\subsection{Training Configuration}

Fine-tuning is performed using SGD with momentum $0.9$, batch size $32$, and initial learning rate $0.01$. Each configuration is trained for $100$ iterations. For each perturbation setting, the resulting encoder is used to compute a corresponding FID variant.

A total of $312$ perturbation configurations are evaluated. Importantly, configuration selection is performed using the training split only. The final SI-FID variant is chosen based on rank-correlation consistency across validation settings, ensuring robustness rather than dataset-specific overfitting.

This procedure adapts the FID framework to stitched-image characteristics through representation calibration, producing a perceptually aligned and noise-aware evaluation metric.

\section{Experiments and Analysis}

This section evaluates the proposed SI-FID through dataset construction, subjective testing, noise-perturbation analysis, comparison with existing indicators, and robustness assessment. Unless otherwise specified, training, validation, and testing data are strictly separated.

\subsection{Dataset}

We employ three types of image sets for training, testing, and subjective evaluation.

\textbf{Training set.}
Training images are taken from the DIR-D dataset~\cite{20}, containing diverse irregular scenes at $512\times384$ resolution.  
A total of 519 base images and 7,266 noise-augmented variants form 7,785 training samples.  
These samples are used exclusively for contrastive fine-tuning and perturbation analysis. No subjective scores from the test sets are used during training or configuration selection.

\textbf{Testing sets.}
Two complementary test sets are constructed, each containing source images, geometrically transformed images, and stitched images generated by four representative stitching algorithms: APAP~\cite{APAP}, SPHP~\cite{SPHP}, ANAP~\cite{ANAP}, and SPW~\cite{SPW}.  
The transformed image is the warped version of the source, and the stitched image is the final composite.

-- \emph{Test Set I}: 20 groups from public benchmark data (80 transformed + 80 stitched images).  

-- \emph{Test Set II}: 20 in-house groups with more diverse textures, parallax, and illumination variations.

Quality assessment focuses on stitched vs. transformed similarity, which reflects the perceptual objective of evaluating compositional integrity after stitching.

\textbf{Subjective evaluation set.}
A subset of 160 stitched images from both test sets is scored by human observers.

\textbf{Why two test sets?}
Test Set~I provides controlled and regular scenes to measure baseline consistency.  
Test Set~II introduces stronger parallax, illumination variation, and structural complexity to evaluate generalization.  
Consistent performance across both sets indicates that the proposed metric does not rely on dataset-specific characteristics.

\begin{figure}[t]
	\centering
	\subfloat[]{\includegraphics[width=0.48\linewidth]{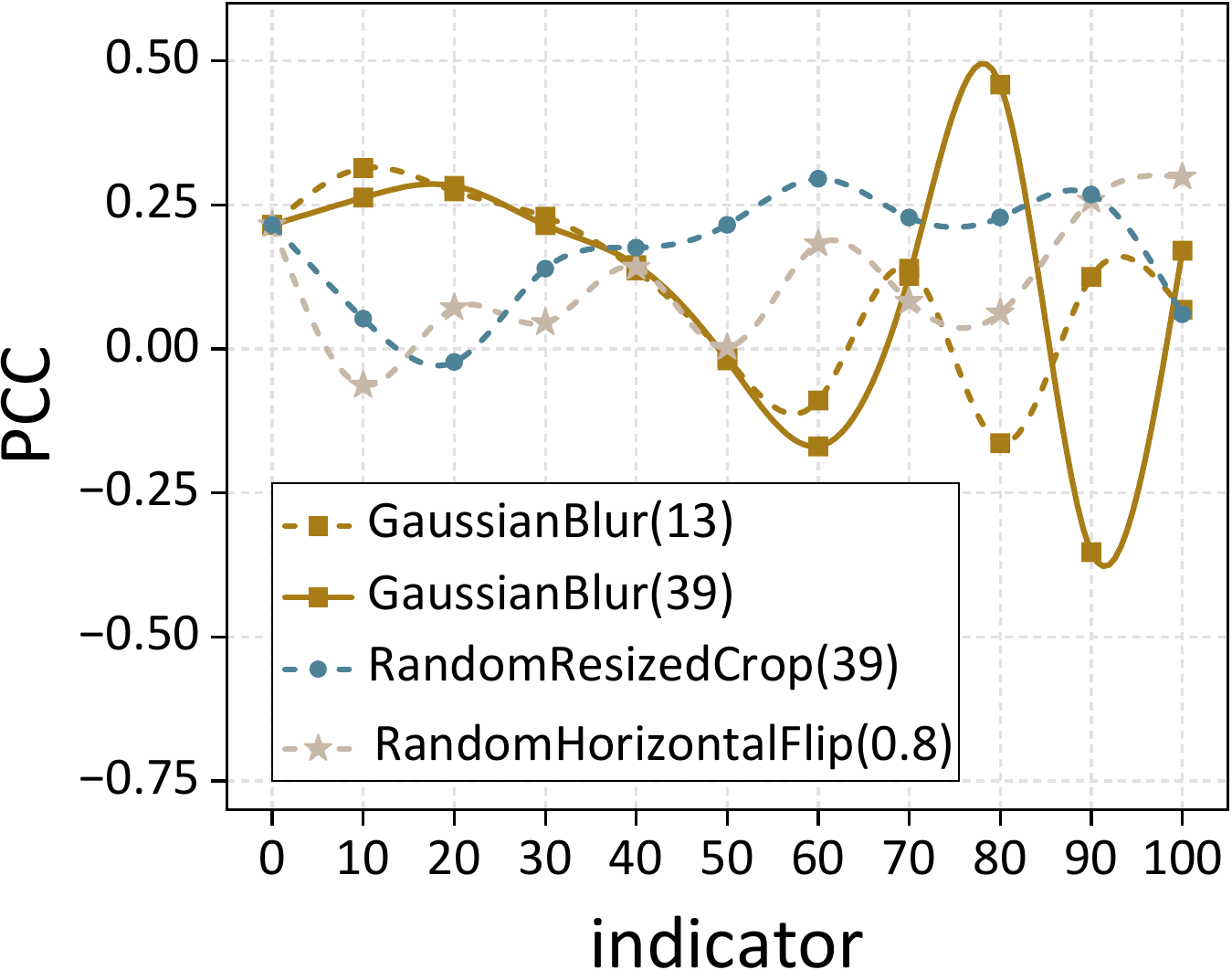}
		\label{fig_6a}}
	\hfill
	\subfloat[]{\includegraphics[width=0.48\linewidth]{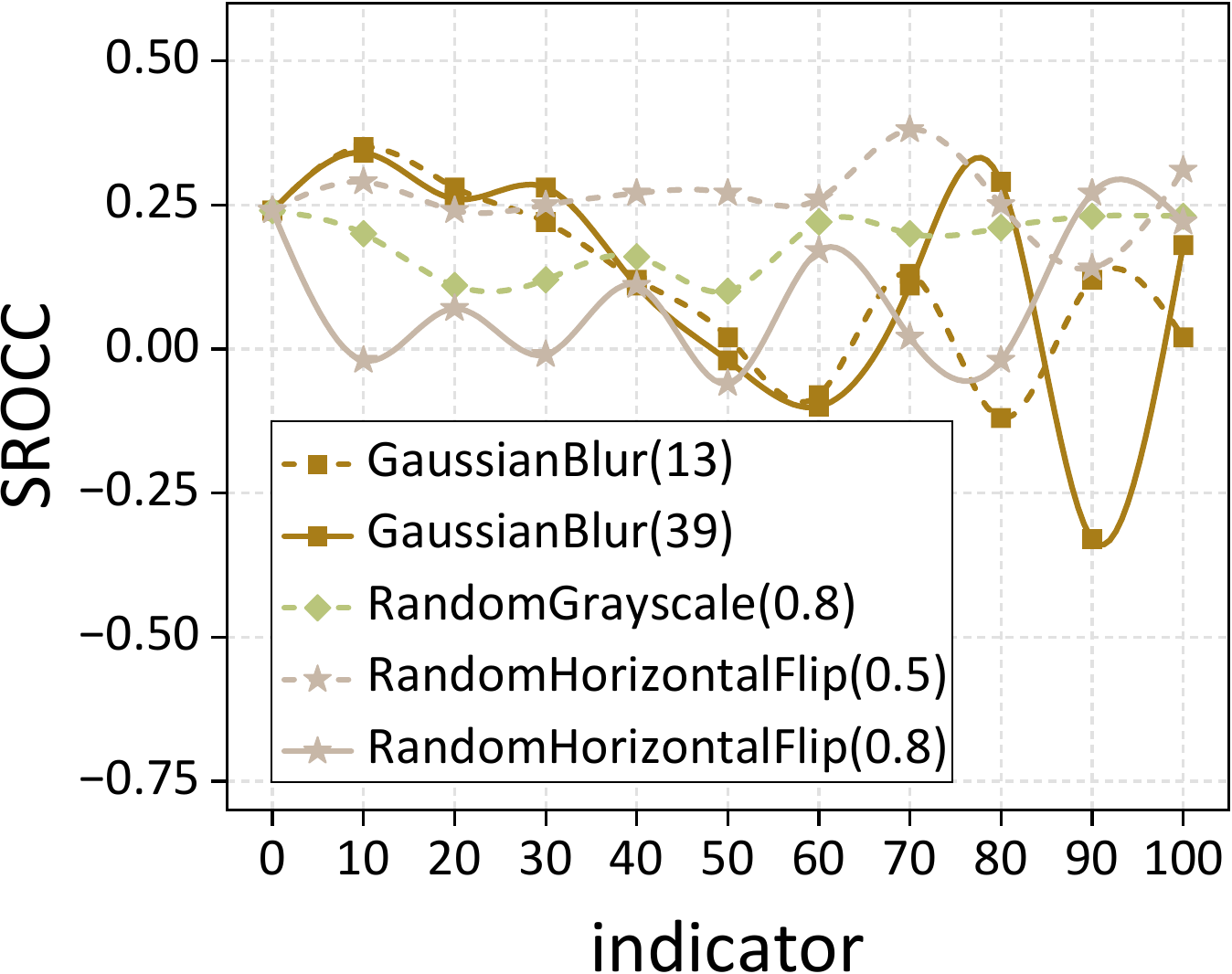}
		\label{fig_6b}}
	
	\vspace{0mm}
	
	\subfloat[]{\includegraphics[width=0.48\linewidth]{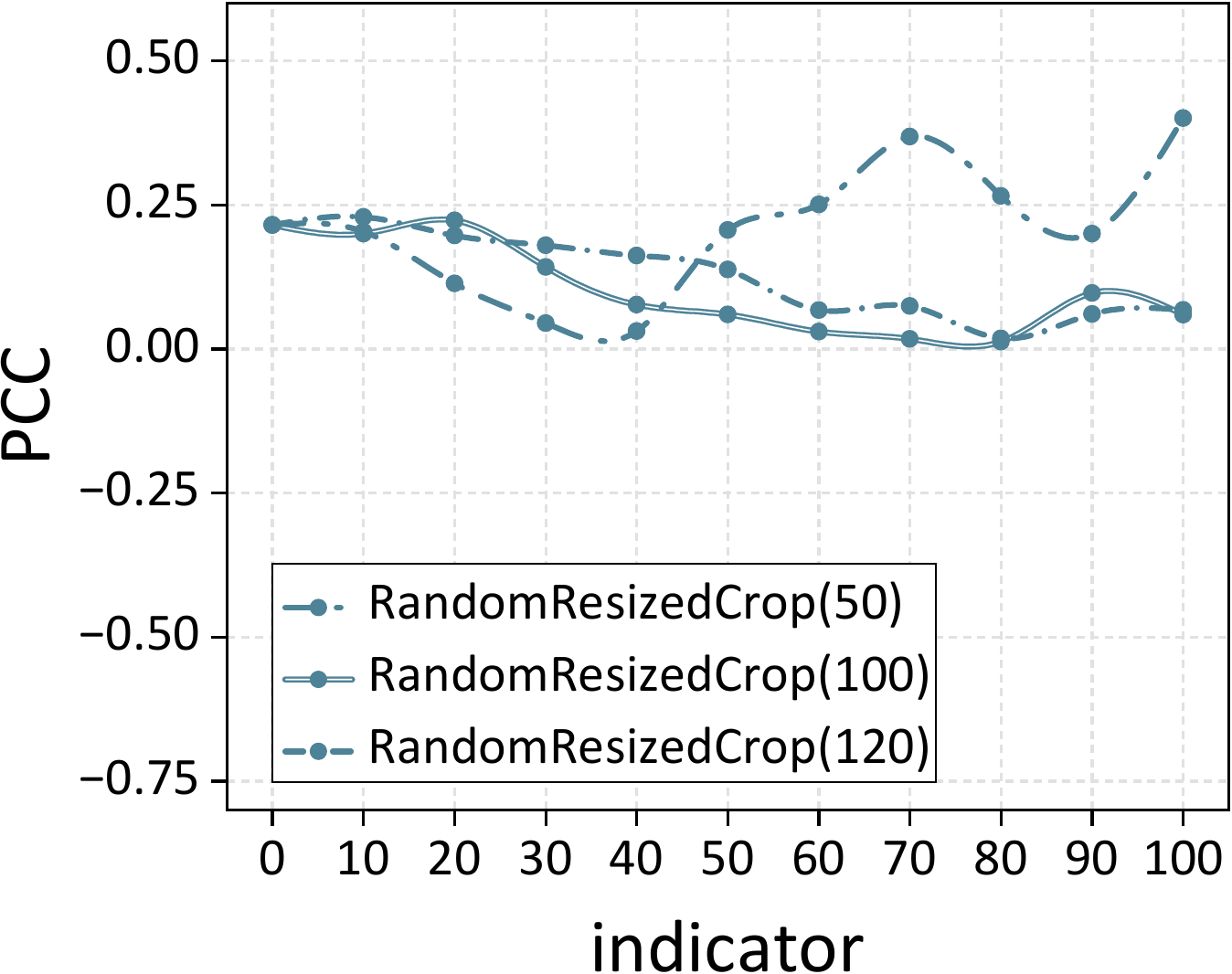}
		\label{fig_6c}}
	\hfill
	\subfloat[]{\includegraphics[width=0.48\linewidth]{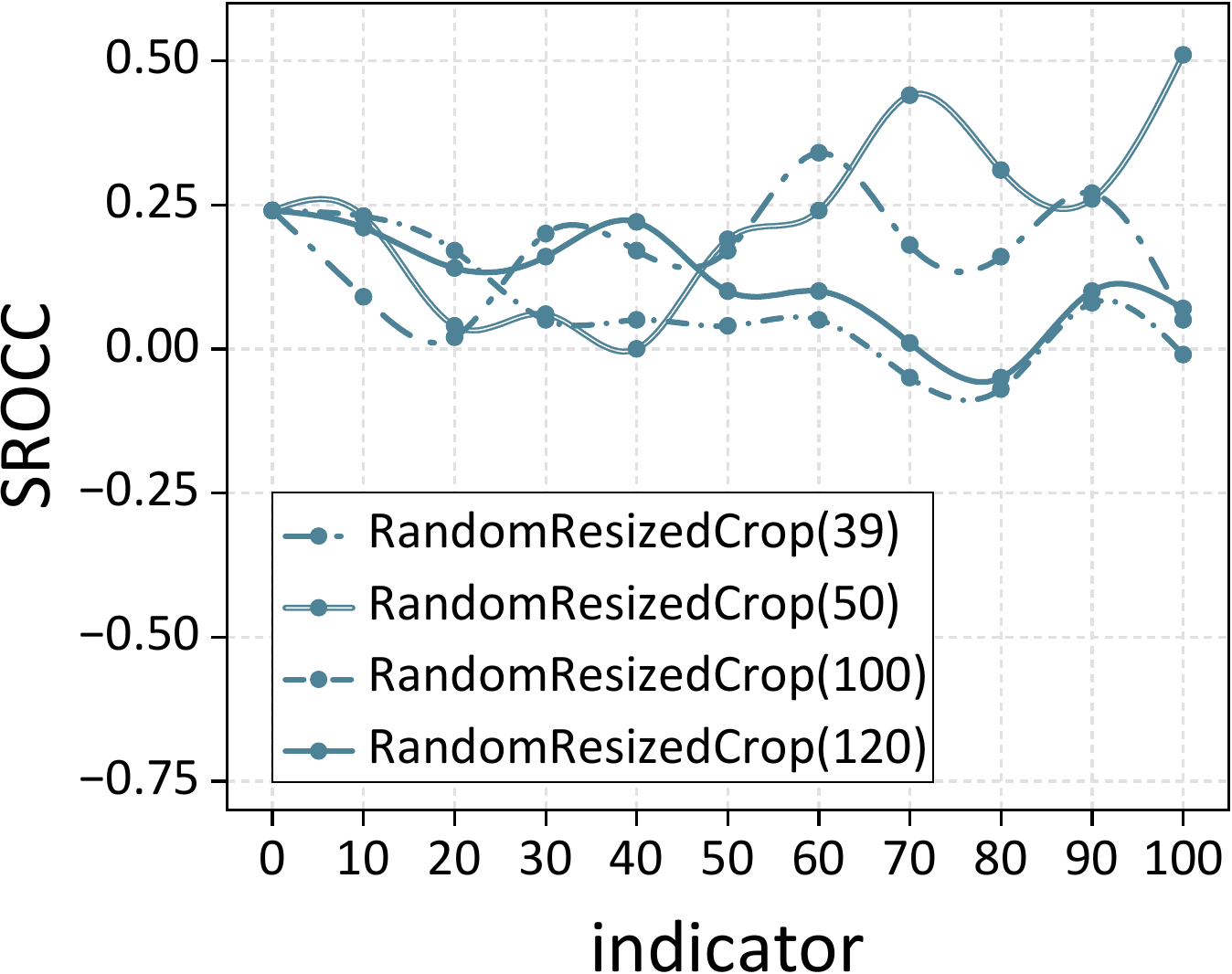}
		\label{fig_6d}}
	
	\vspace{0mm}
	
	\subfloat[]{\includegraphics[width=0.48\linewidth]{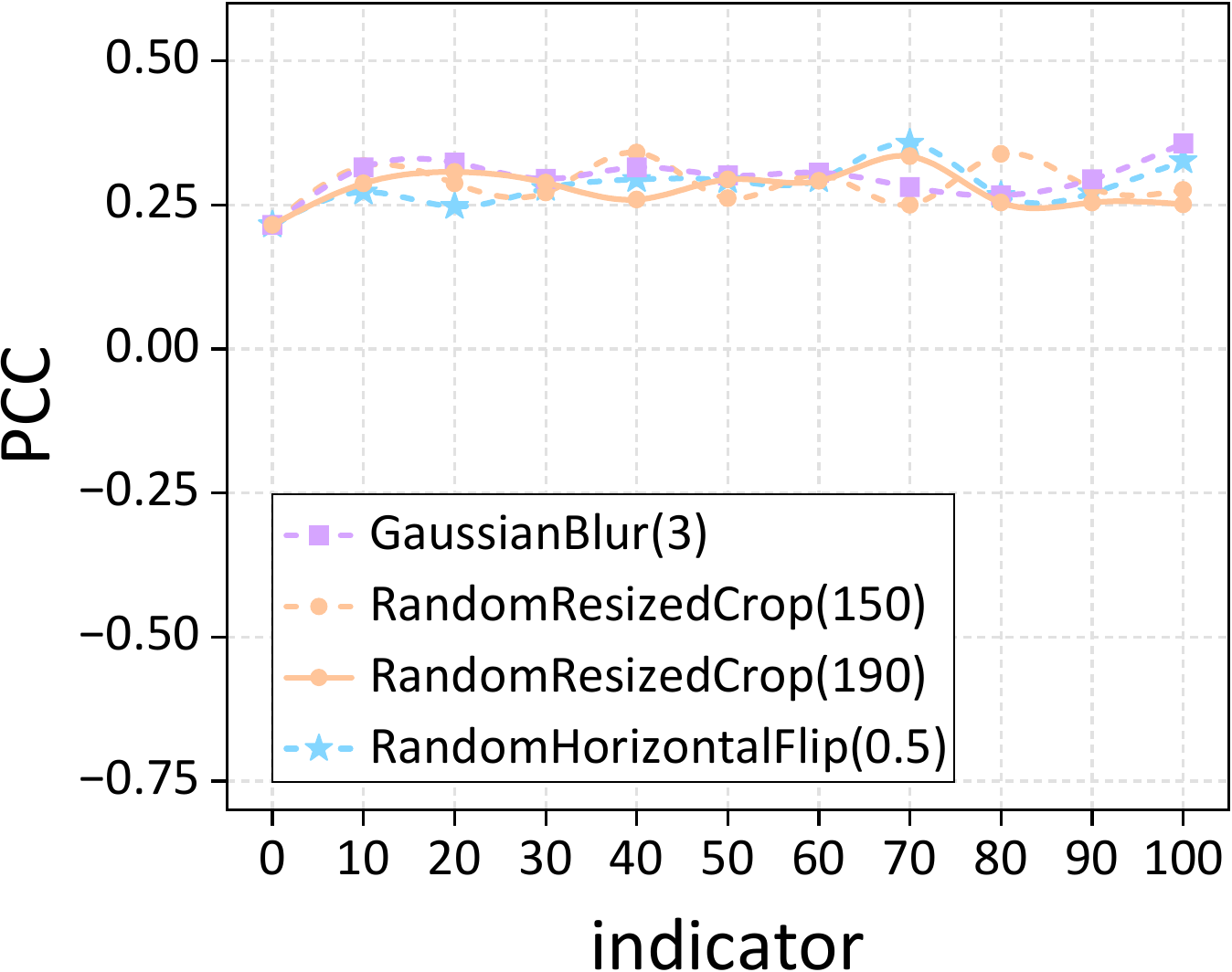}
		\label{fig_6e}}
	\hfill
	\subfloat[]{\includegraphics[width=0.48\linewidth]{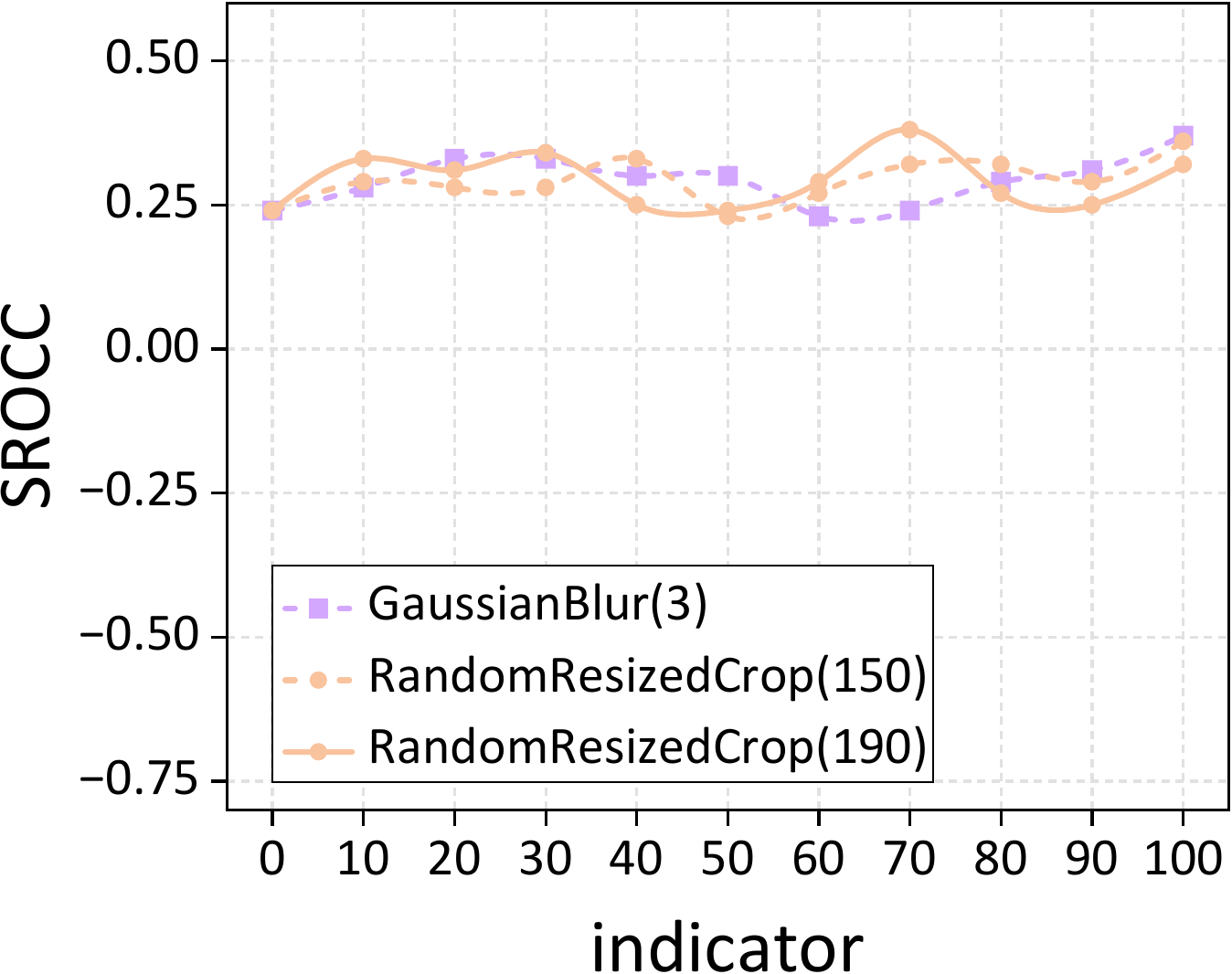}
		\label{fig_6f}}
	
	\vspace{0mm}
	
	\subfloat[]{\includegraphics[width=0.48\linewidth]{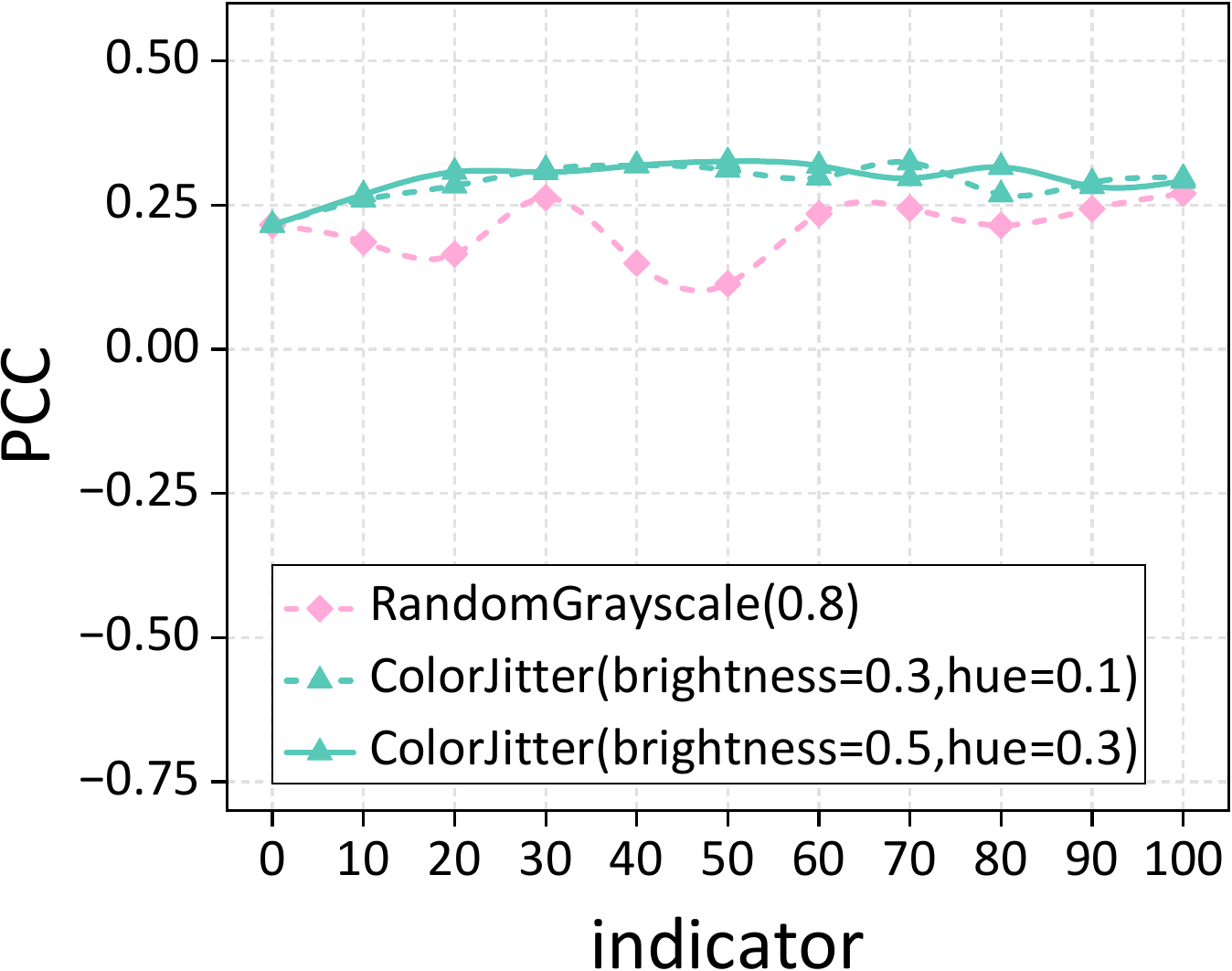}
		\label{fig_6g}}
	\hfill
	\subfloat[]{\includegraphics[width=0.48\linewidth]{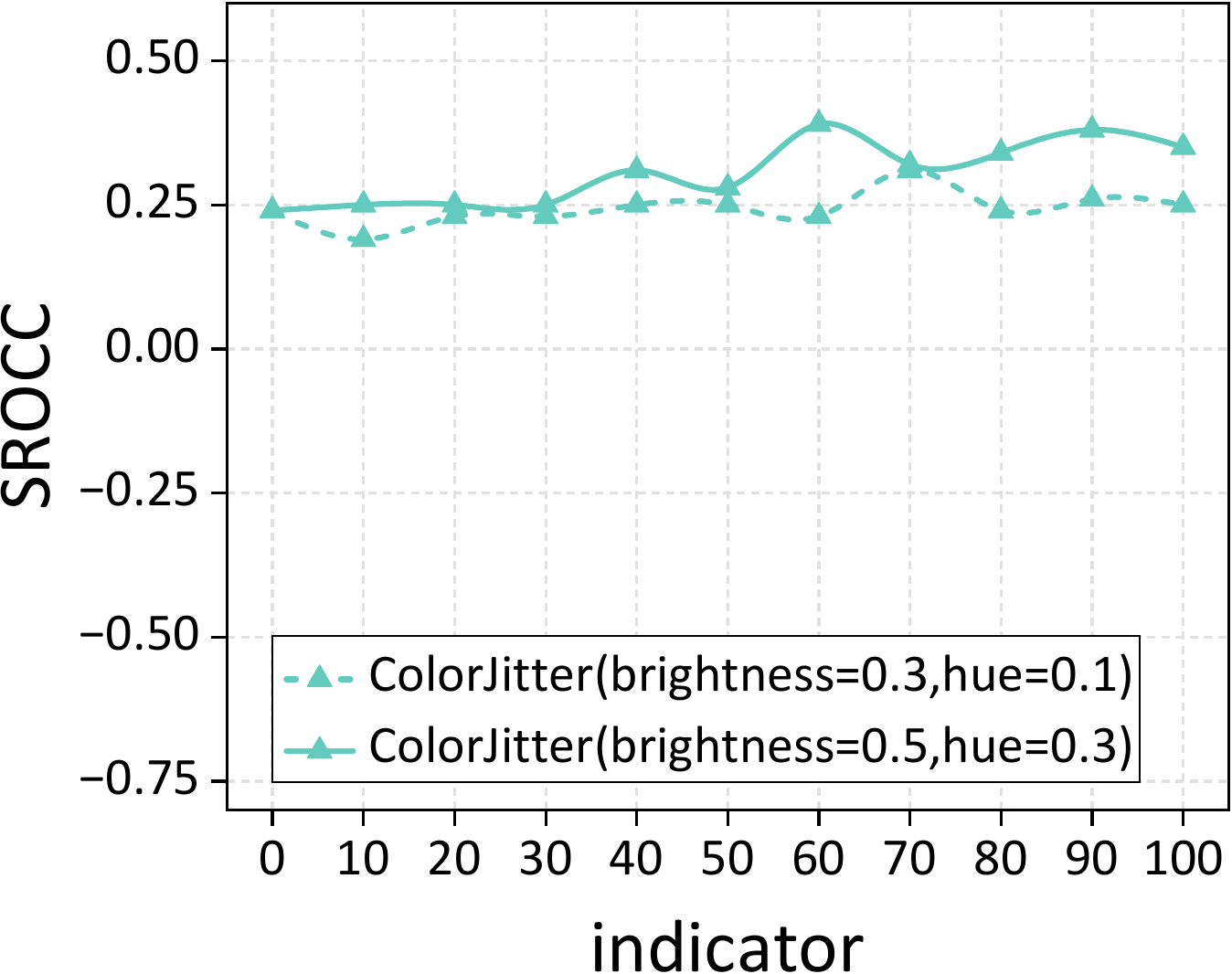}
		\label{fig_6h}}
	\caption{
		Rank-correlation trends under different noise perturbations. 
		Each subplot corresponds to one augmentation strategy used as a controlled noise source. 
		(a)--(d) depict negative or excessive perturbations that yield unstable or degraded correlations. 
		(e)--(h) show moderate perturbations (e.g., Gaussian blur, ColorJitter, RandomResizedCrop) that produce smoother upward trends and better agreement with human ratings.}
	\label{fig:noise_curves}
\end{figure}

\begin{figure}[t]
	\setlength{\abovecaptionskip}{-0.2cm}
	\begin{center}
		{
			\label{fig_7}
			\centering
			\includegraphics[width=0.455\linewidth]{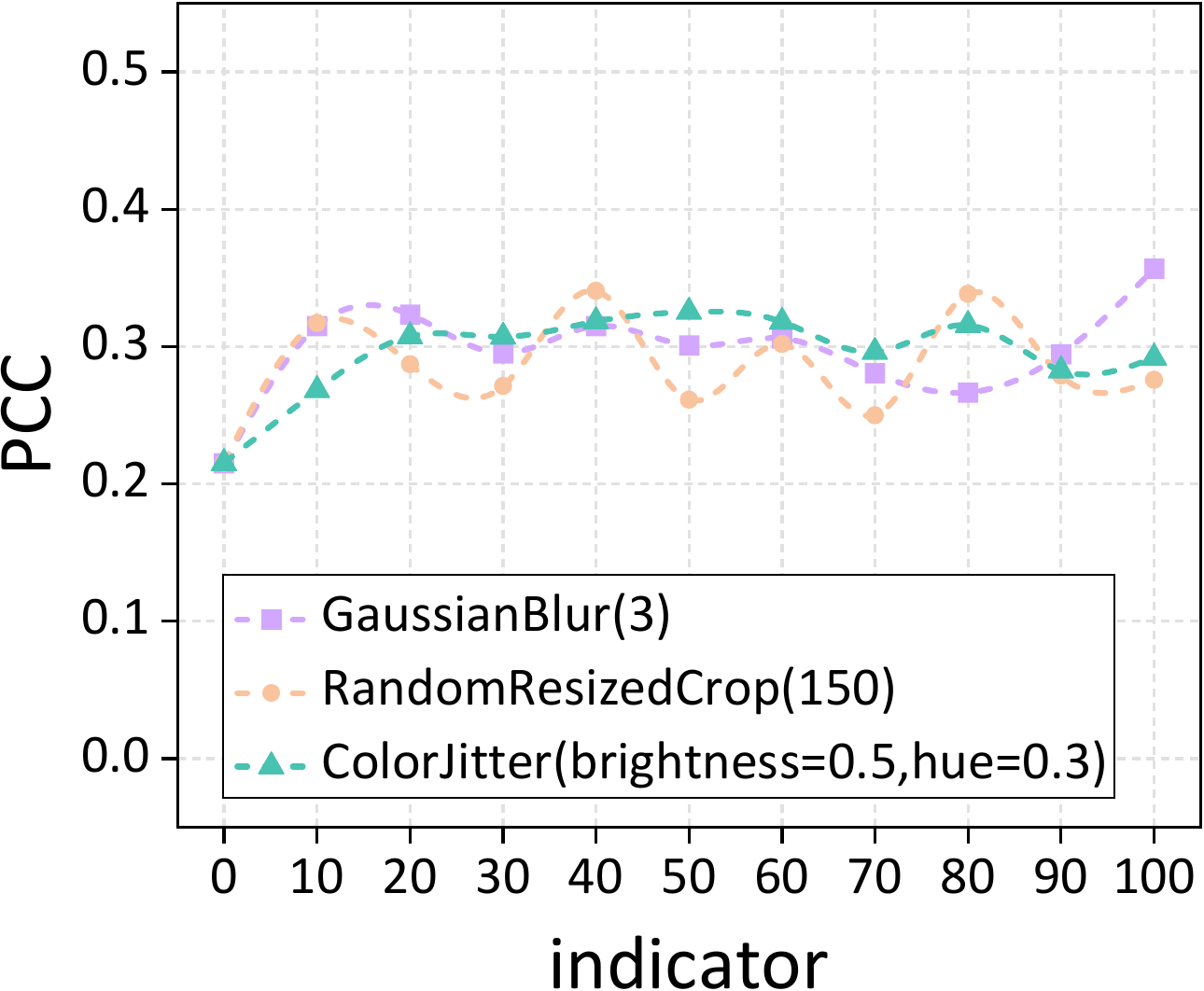}
		}\hspace{-2mm}
		{
			\centering
			\includegraphics[width=0.455\linewidth]{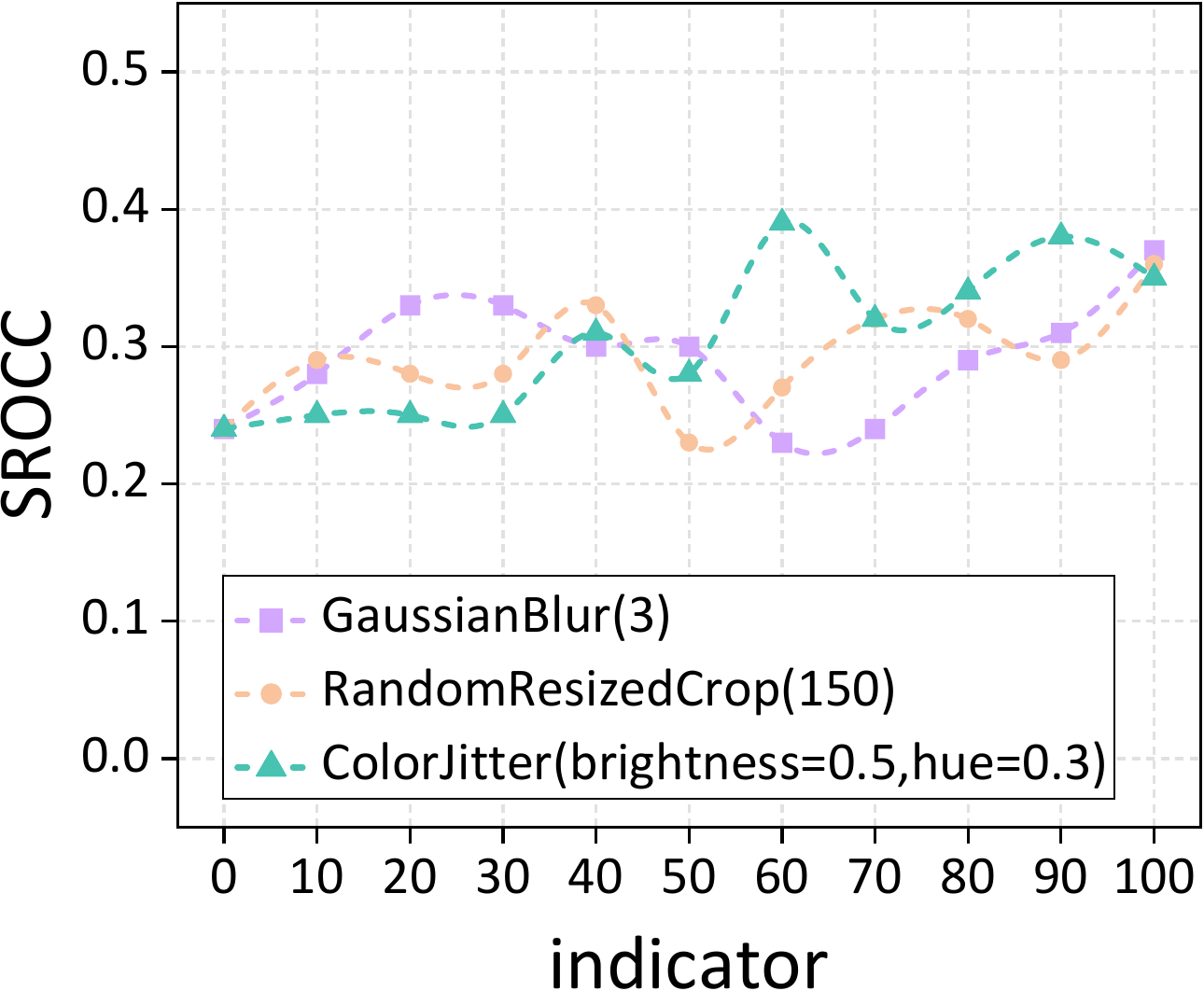}
		}
	\end{center}
	\caption{
		Focused comparison of positive perturbations.
		Among the evaluated settings, ColorJitter (brightness = 0.5, hue = 0.3) delivers the most stable and consistent improvement in FID–human rank correlation.
	}
	\label{fig:positive_noise}
\end{figure}

\subsection{Performance Assessment}

\subsubsection{Subjective Evaluation}

Fourteen experienced observers rated each stitched image on a 0–100 scale with respect to color fidelity, sharpness, and visible stitching artifacts.  
To reduce individual scoring bias, the scores from each evaluator were standardized before aggregation:

\begin{equation}
x_i'=\frac{x_i-\mu_i}{\delta_i}, \qquad 
S=\frac{1}{n}\sum_{i=1}^{n}x_i',
\end{equation}

where $x_i$ is the raw score assigned by the $i$-th evaluator, $\mu_i$ and $\delta_i$ denote that evaluator’s mean and standard deviation across all rated images, $n=14$, and $S$ represents the final aggregated subjective quality score.

\subsubsection{Effect of Noise Perturbations}

Perceptual consistency is quantified using Pearson (PCC) and Spearman (SROCC) correlations:

\begin{align}
r_p &= \frac{\operatorname{cov}(X,Y)}{\sigma(X)\sigma(Y)}, \\
r_s &= 1-\frac{6\sum_i d_i^2}{n(n^2-1)}.
\end{align}

Fig.~\ref{fig:noise_curves} presents correlation trends across 100 representative perturbation configurations (index 0 corresponds to the original FID).  
Each configuration is evaluated independently, and correlation behavior is examined on both test sets to analyze stability rather than isolated peak performance.

In Fig.~\ref{fig_6a}–\ref{fig_6d}, excessive or negative perturbations lead to unstable or decreasing trends, indicating that overly strong distortions disrupt semantic consistency.  
In contrast, Fig.~\ref{fig_6e}–\ref{fig_6h} show that moderate perturbations—Gaussian Blur (3), ColorJitter (0.5, 0.3), and RandomResizedCrop (150)—yield smoother and consistently upward trends in both PCC and SROCC across datasets.

Fig.~\ref{fig:positive_noise} compares the three strongest-performing perturbations.  
Among them, ColorJitter (0.5, 0.3) demonstrates the most stable and monotonic correlation improvement on both test sets.  
The final SI-FID configuration is selected based on cross-dataset consistency and trend stability rather than a single maximum correlation value.

\subsubsection{Comparison with Representative IQA Metrics}

We compare SI-FID with widely used full-reference (FR) metrics (MSE, PSNR, SSIM, FSIM, LPIPS, FID) and no-reference (NR) metrics (NIQE, BRISQUE).  
All objective scores are normalized for consistent comparison.

\begin{figure}[h]
	\setlength{\abovecaptionskip}{-0.2cm}
		\centering
		\subfloat[]{\includegraphics[width=0.48\linewidth]{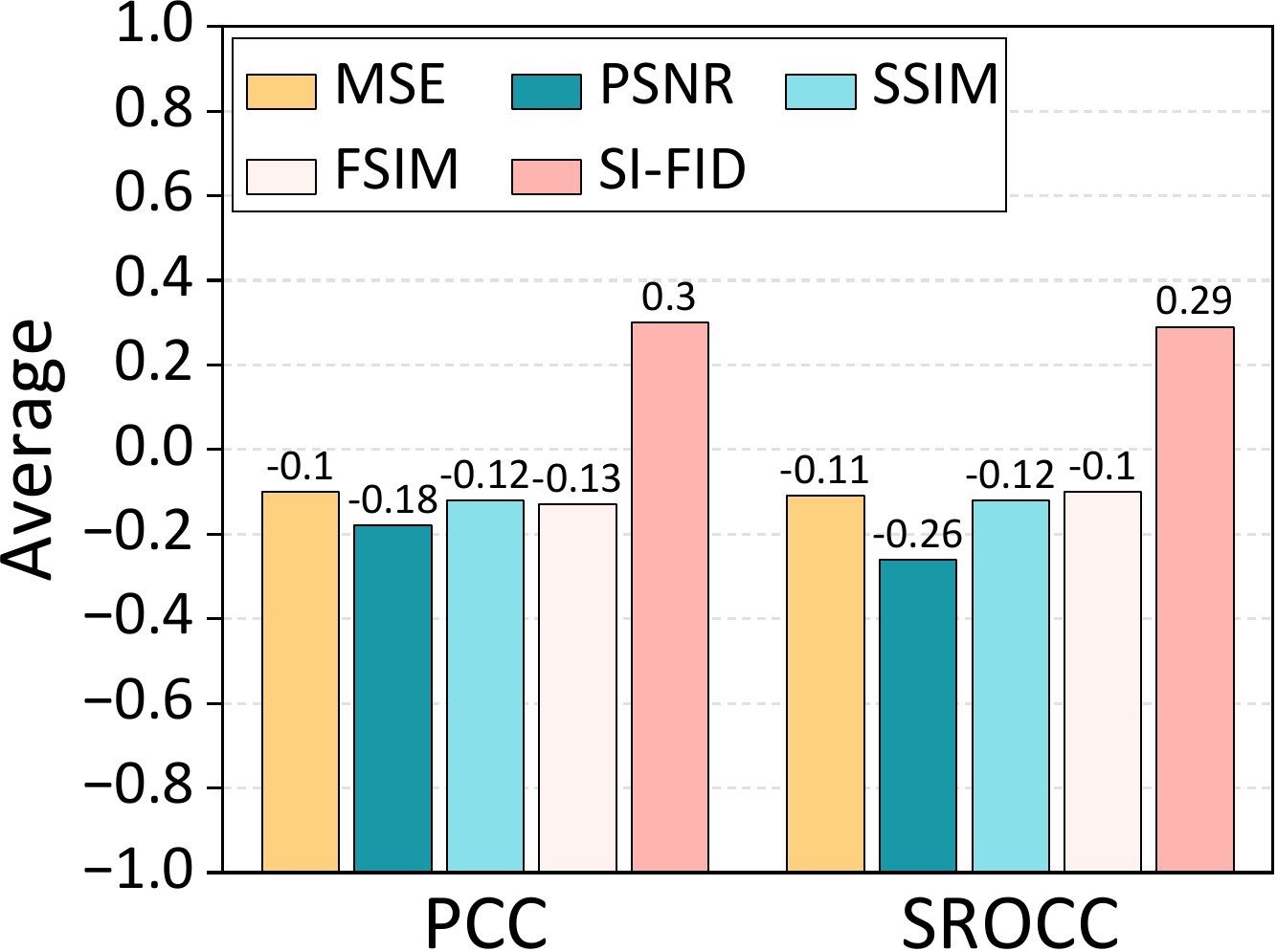}
			\label{fig_8a}}
		\hfill
		\subfloat[]{\includegraphics[width=0.48\linewidth]{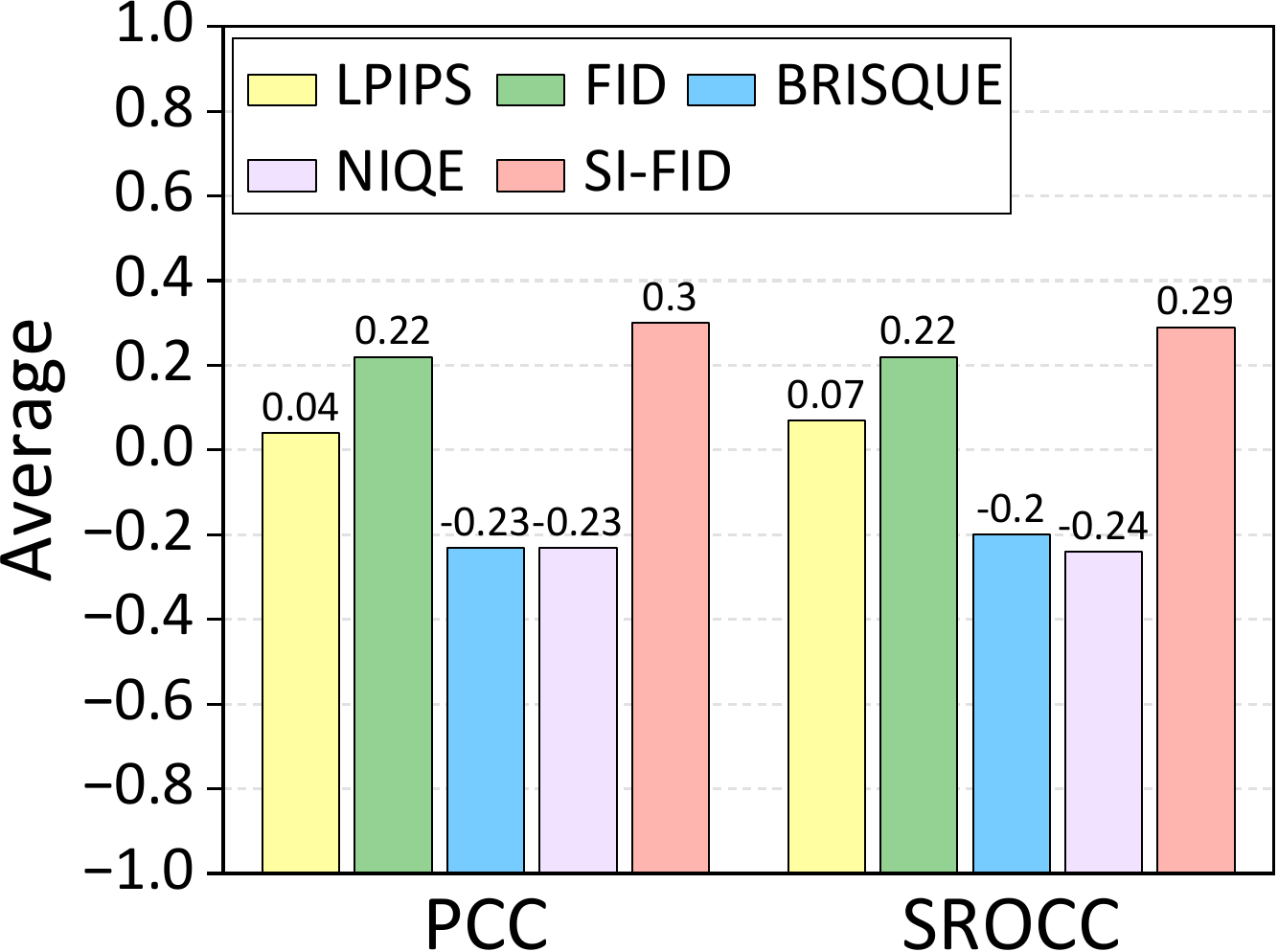}
			\label{fig_8b}}
		
		\vspace{-3mm}
		
		\subfloat[]{\includegraphics[width=0.48\linewidth]{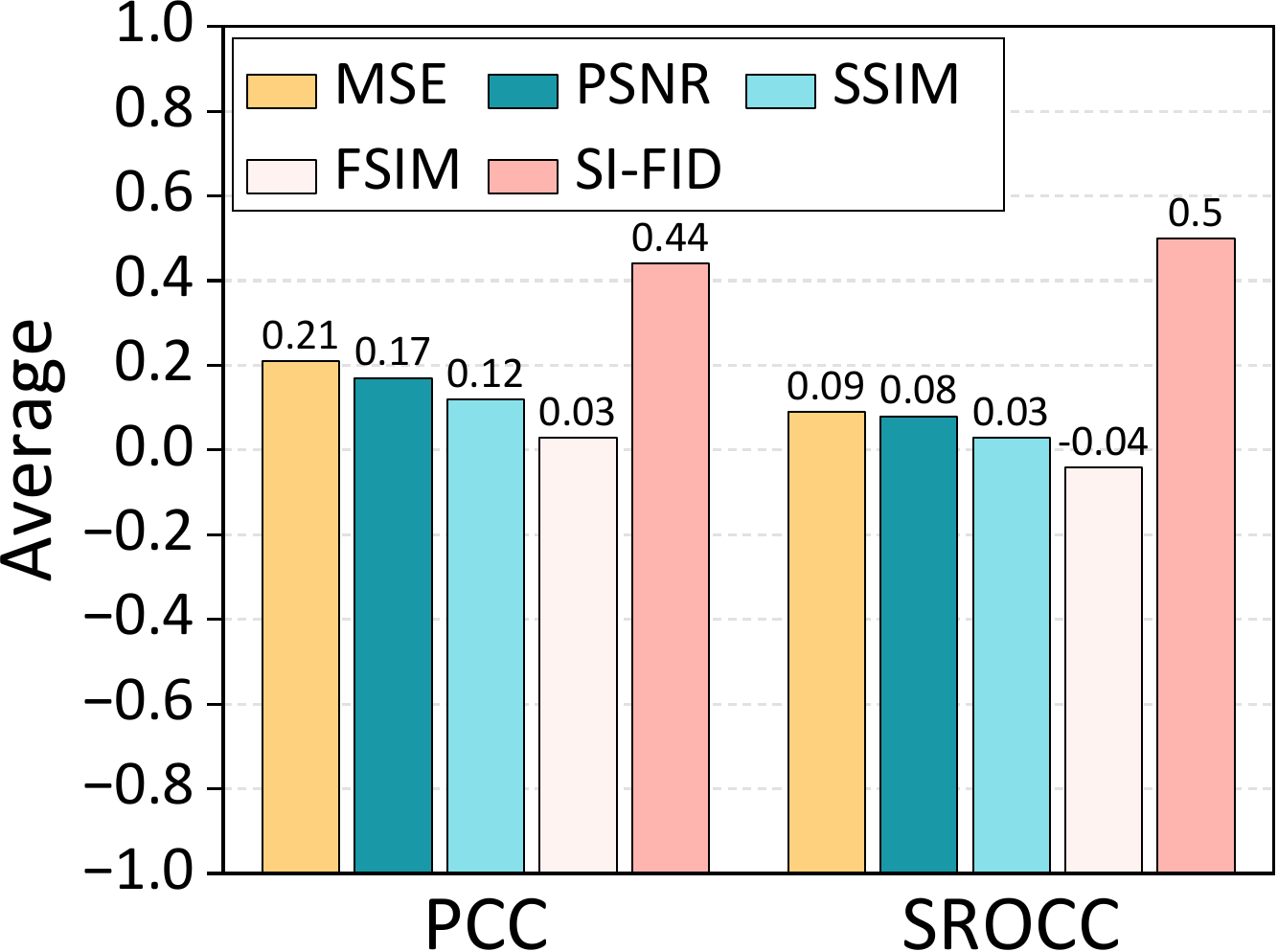}
			\label{fig_8c}}
		\hfill
		\subfloat[]{\includegraphics[width=0.48\linewidth]{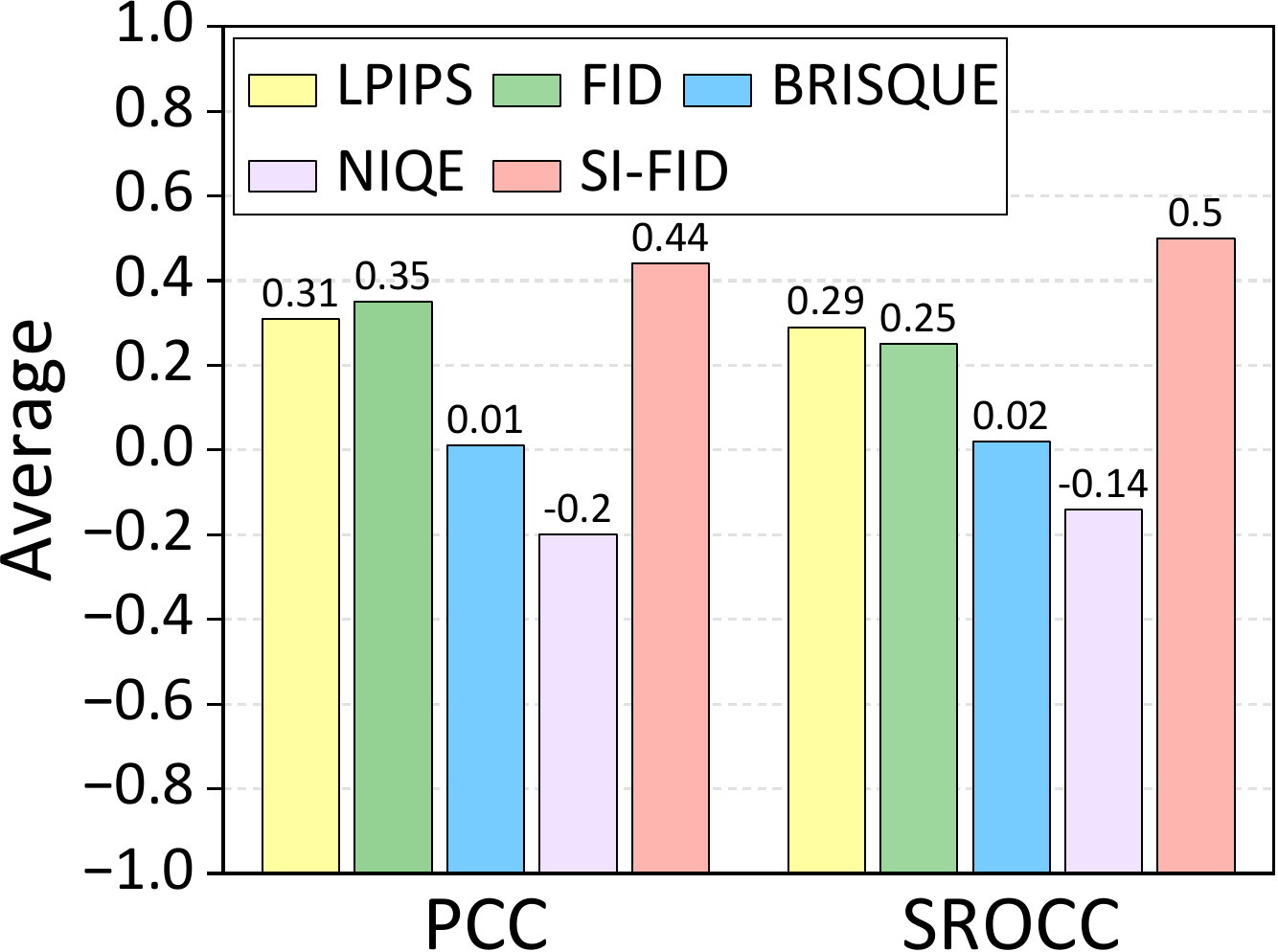}
			\label{fig_8d}}
		\vspace{4mm}
	\caption{
		Average rank-correlation coefficients (PCC and SROCC) of representative IQA metrics on Test Set~I and Test Set~II. 
		Subplots (a,b) show results on Test Set~I, and (c,d) on Test Set~II. 
		Classical full-reference (FR) metrics (MSE, PSNR, SSIM, FSIM), learning-based metrics (LPIPS, FID), no-reference (NR) metrics (NIQE, BRISQUE), and SI-FID are compared. 
		SI-FID consistently achieves the highest correlation, indicating superior perceptual alignment across both datasets.}

	\label{fig:rank_corr}
\end{figure}

Fig.~\ref{fig:rank_corr} summarizes the average PCC and SROCC across both test sets.  
Subplots (a,c) report classical FR metrics, while (b,d) present learning-based and NR indicators.  
Across all categories and both datasets, SI-FID achieves the highest average correlation with human judgments, demonstrating consistent improvement over conventional FR and NR measures.

\subsubsection{Robustness and Stability Analysis}

\begin{figure}[h]
	\setlength{\abovecaptionskip}{-0.2cm}
	\begin{center}
		\centering
		\subfloat[]{\includegraphics[width=0.48\linewidth]{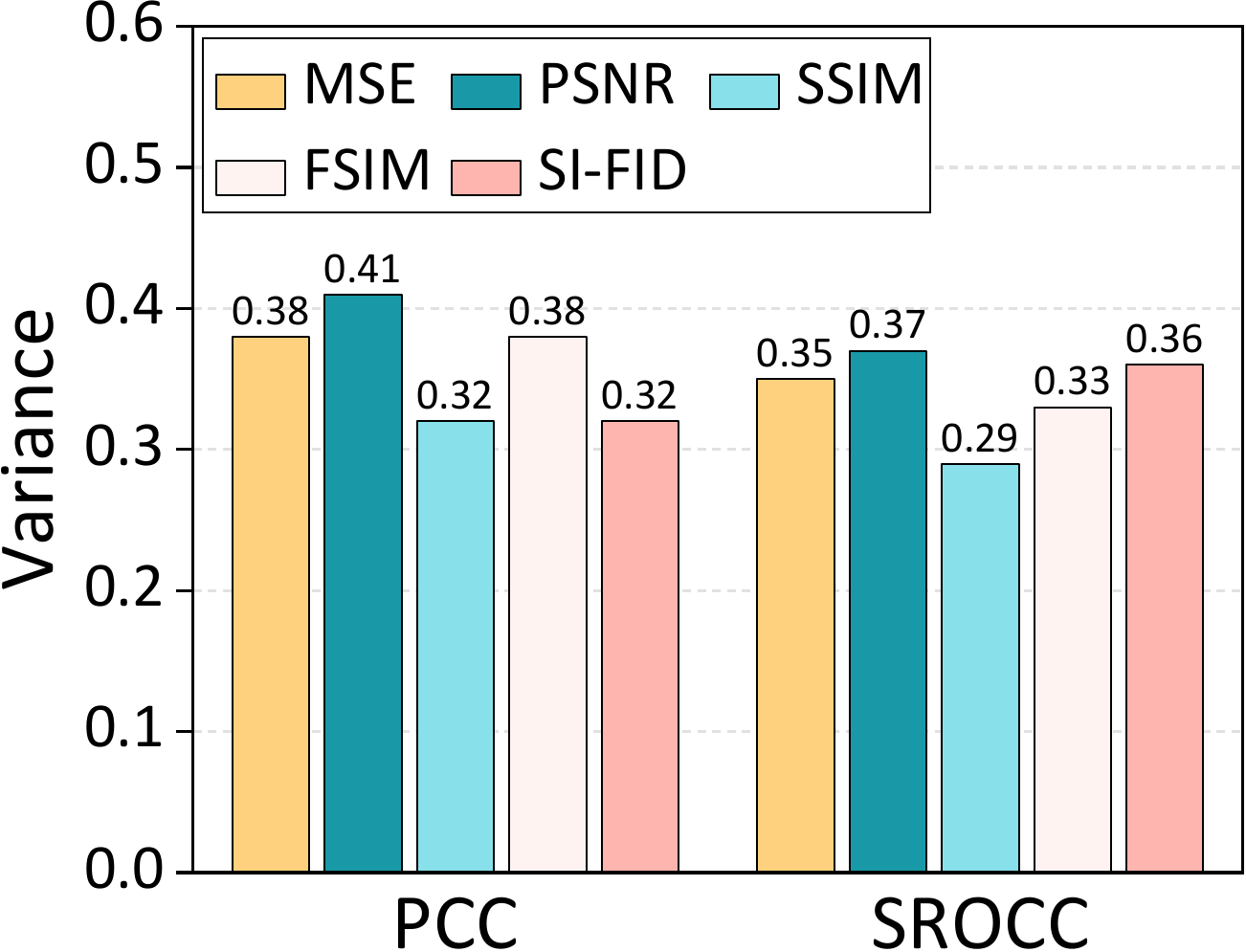}
			\label{fig_9a}}
		\hfill
		\subfloat[]{\includegraphics[width=0.48\linewidth]{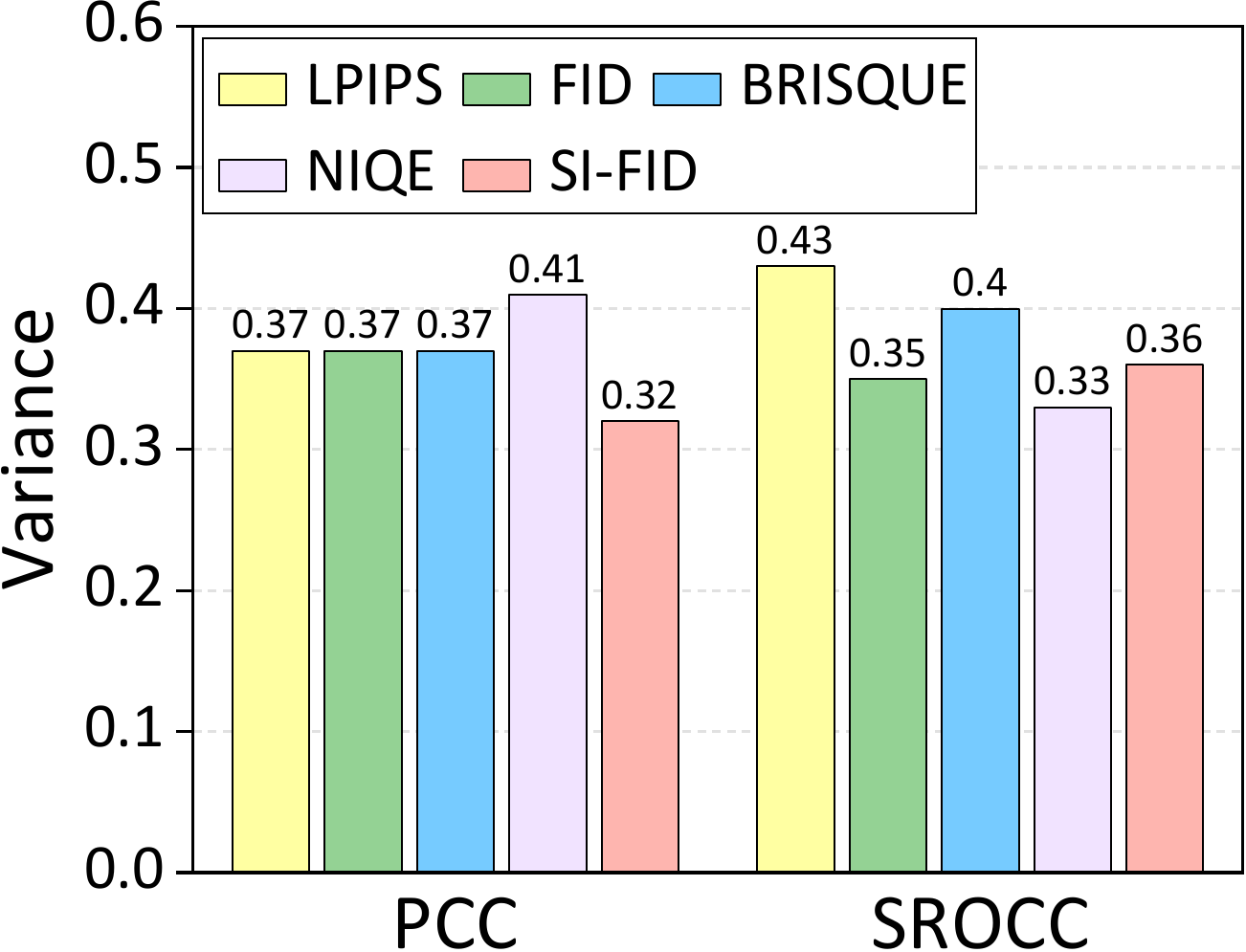}
			\label{fig_9b}}
		
		\vspace{-4mm}
		
		\subfloat[]{\includegraphics[width=0.48\linewidth]{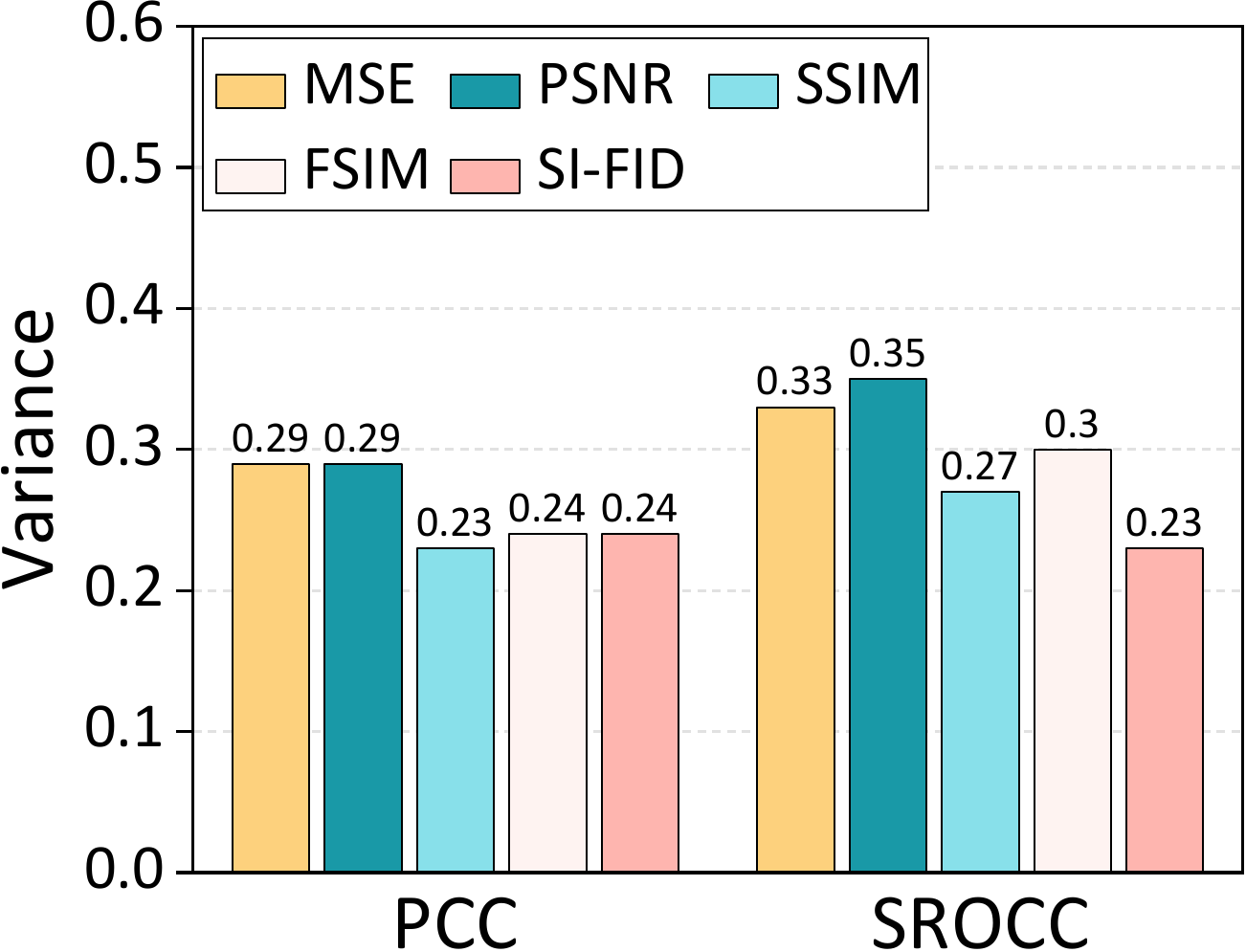}
			\label{fig_9c}}
		\hfill
		\subfloat[]{\includegraphics[width=0.48\linewidth]{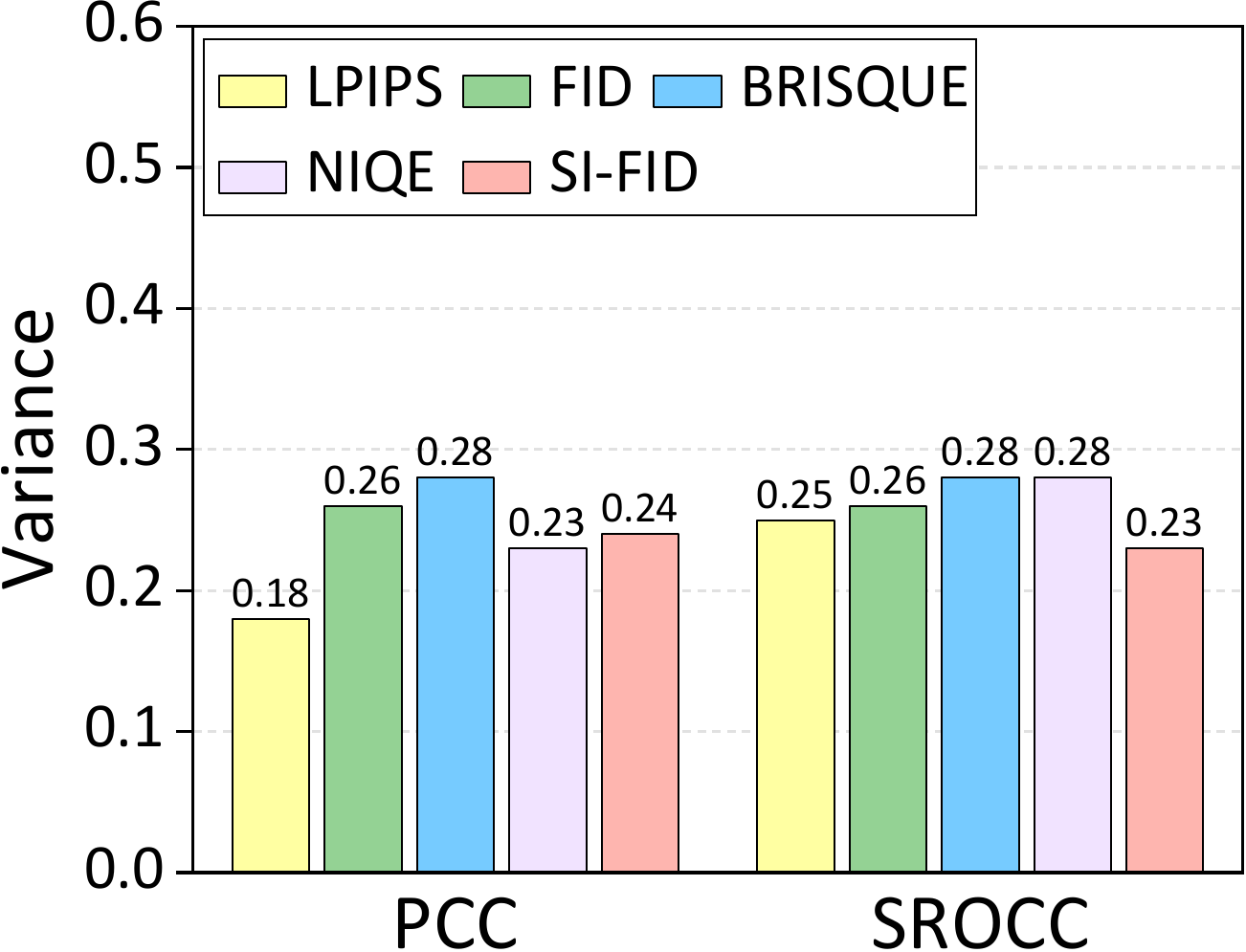}
			\label{fig_9d}}
		\vspace{4mm}
	\end{center}
	\caption{
		Variance of PCC and SROCC across different IQA metrics. 
		Subplots (a,b) report results on Test Set~I, and (c,d) on Test Set~II. 
		(a,c) compare traditional FR metrics with SI-FID; 
		(b,d) include learning-based and NR metrics. 
		Lower variance indicates greater stability, and SI-FID demonstrates the lowest or near-lowest variance across both datasets, confirming its robustness and generalizability.
	}
	
	\label{fig:rank_var}
\end{figure}

Robustness is evaluated via the variance of PCC and SROCC across test sets, as shown in Fig.~\ref{fig:rank_var}.  
Subplots (a,c) correspond to FR metrics, and (b,d) correspond to learning-based and NR metrics.

In all cases, SI-FID exhibits the lowest or near-lowest variance, indicating stable behavior across different scene structures and distortion patterns.  
On Test Set~I, SI-FID slightly outperforms the average stability of FR metrics; on Test Set~II, it remains among the most robust indicators.  
These results demonstrate that SI-FID not only improves perceptual correlation but also maintains strong consistency under varied testing conditions.

\section{Conclusion}

This work presented SI-FID, a noise-aware extension of the Fréchet Inception Distance tailored for stitched image quality assessment. By introducing controlled perturbations through data augmentation and fine-tuning InceptionV3 within a contrastive framework, SI-FID enhances representation sensitivity to stitching-induced artifacts such as misalignment and ghosting—distortions that conventional objective metrics often fail to capture reliably.

Comprehensive experiments on two complementary test sets demonstrate that SI-FID achieves higher correlation with human subjective judgments and exhibits stable performance across diverse scene conditions. These results indicate that perturbation-driven representation calibration is an effective strategy for improving perceptual alignment in feature-distance–based evaluation.

While SI-FID is designed for stitched images, the proposed noise-aware adaptation framework can potentially be extended to other distortion-sensitive perceptual assessment tasks. Future work will investigate adaptive perturbation strategies, alternative backbone architectures, and extensions to video stitching and dynamic panoramic scenarios. Overall, SI-FID provides a practical and perceptually consistent metric for stitched image evaluation.

\bibliographystyle{IEEEbib}
\bibliography{fid_bibliography}

\end{document}